\documentclass[12pt]{report}
\usepackage{amstex,amssymb}

\renewcommand{\theequation}{\arabic{equation}}

\def\ds{\displaystyle}

\def\aam{A^{a}_{\mu}}

\def\pmu{\partial_{\mu}}
\def\pnu{\partial_{\nu}}
\def\fmn{F_{\mu\nu}}

\def\eabc{\epsilon_{abc}}

\def\ra{\rightarrow}

\begin{document}

\title{Nontopological Methods \\ for Determining Topological Charge \\ for Bosons and Fermions in Flat Spacetime: \\ }
\author{Joseph Saaty, PhD \\
The Union Institutute \\
School of Interdisciplinary Arts and Sciences} 

\maketitle

\pagenumbering{arabic}
\pagestyle{plain}
\baselinestretch{2}
\baselineskip=22pt

\def\theequation{\arabic{equation}}
\setcounter{equation}{0}

\begin{abstract}
An alternative method to the topological instanton solution for deriving 
an expression for the topological charge is presented.  
This alternative method involves the use of relativistic quantum field 
theory and covariant electrodynamics.  In the case of bosons, this 
method is consistent with the instanton solution in predicting that 
topological charge is quantized. But furthermore, 
this method led to the new results that topological charge for
fermions cannot be quantized, whereas the instanton solution cannot
distinguish between bosons (quantized) and fermions (not quantized). Thus
the new technique produced results that were previously unobtainable.
Mathematics Subject Classifications (1991): //
\end{abstract}


\noindent

To prove that the topological charge
$ Q = \dfrac{1}{8\pi} \ds\int \epsilon_{\mu\nu} 
\hat{\phi}\cdot(\pmu\hat{\phi} \times \pnu\hat{\phi}) \ d^3x$
is quantized \cite{12} by using covariant electrodynamics and relativistic quantun field theory, we proceed as follows:

\bigskip

\noindent{\bf Proof:}
$\ds  \fmn   = \pmu A_{\nu} - \pnu A_{\mu} - \frac{1}{e|\phi|^3} 
                                    \eabc \phi^a(\pmu\phi^b)(\pnu\phi^c)$, 
\begin{align*}
   A_{\mu} &= \frac{1}{|\phi|} \phi^a\aam \ \ \mbox{and} \ \ 
                                     \hat{\phi} = \frac{\phi}{|\phi|}\\ 
  K^{\mu} &= \pnu \tilde{F}^{\mu\nu} \ \ \mbox{is the magnetic current} \\ 
\tilde{F}^{\mu\nu} &= \frac 12 \epsilon^{\mu\nu\sigma\rho} F^{\sigma\rho} 
  \qquad \mbox{so that,}\\
K^{\mu} &= -\frac{1}{2e} \epsilon^{\mu\nu\rho\sigma}  
                             \eabc \pnu\hat{\phi}^a \partial_{\sigma} \hat{\phi}^c
                             \partial_{\rho} \hat{\phi}^b \\ 
 K^{\mu} &= \frac 12 \epsilon^{\mu\nu\sigma\rho}\pnu F^{\sigma\rho}\\ 
 K^{o} &= \frac 12 \epsilon^{\nu\sigma\rho}\pnu F^{\sigma\rho}   
\end{align*}
Since $\pmu K^{\mu} = 0$, the conserved magnetic charge can be written as
$M = \dfrac{1}{4\pi}\ds\int K^od^3x$.
\begin{align*}
 M &= -\frac{1}{8e\pi} \int \epsilon^{ijk}\eabc \partial_i \hat{\phi}^a 
             \partial_j \hat{\phi}^b \partial_k \hat{\phi}^c \ d^3x\\ 
 M &= \frac{1}{8e\pi} \oint \epsilon^{ijk}\eabc \hat{\phi}^a \partial_j 
           \hat{\phi}^b \partial_k \hat{\phi}^c \ dS_i 
\end{align*}
i.e., 
\begin{equation}\label{eq6}
   K^o = -\dfrac{1}{2e}  \epsilon^{ijk}\eabc \partial_i \hat{\phi}^a 
        \partial_j \hat{\phi}^b \partial_k \hat{\phi}^c 
\end{equation}
but, $\ds K^{\mu} = -\dfrac{1}{2e} \epsilon^{\mu\nu\rho\sigma} \eabc 
\pnu\hat{\phi}^a \partial_{\rho} \hat{\phi}^b \partial_{\sigma}\hat{\phi}^c$.
If \ \ $i = \mu $, \ \  $j = \nu$, \ \ $k = \rho$
\begin{equation}\label{eq7}
\mbox{then we will get}  \quad  K^{\mu} = -\frac{1}{2e} \epsilon^{ijk\sigma}\eabc\partial_i 
      \hat{\phi}^a \partial_j \hat{\phi}^b \partial_k \hat{\phi}^c
\end{equation}
\medskip         

\noindent Maxwell's equation in covariant form is $\pmu F^{\mu\nu} = j^{\nu}$.
\begin{align*}
B_i &= \frac 12 \epsilon_{ijk} F^{jk}, \ \ 
                        \vec{B} \ \ \mbox{is the magnetic field}\\ 
\nabla\cdot\vec{B} &= \partial_i B_i \ \mbox{so that} \ 
              \nabla\cdot\vec{B} = \frac 12 \epsilon_{ijk}  \partial_i F^{jk}\\
K^{\mu} &= \pnu \tilde{F}^{\mu\nu}, \ \ K^o = \pnu \tilde{F}^{o\nu}, \ \
                    B^i = -\frac 12 \epsilon^{ijk} F_{jk} \\ 
K^{\mu} &= -\frac{1}{2e} \epsilon^{\mu\nu\rho\sigma} \eabc 
     \pnu\hat{\phi}^a \partial_{\rho} \hat{\phi}^b \partial_{\sigma}\hat{\phi}^c\\ 
\pnu \tilde{F}^{\mu\nu}&= -\frac{1}{2e} \epsilon^{\mu\nu\rho\sigma} \eabc 
     \pnu\hat{\phi}^a \partial_{\rho} \hat{\phi}^b \partial_{\sigma}\hat{\phi}^c\\ 
K^o &= -\frac{1}{2e} \epsilon^{ijk} \eabc 
                \partial_i \hat{\phi}^a  
                \partial_j \hat{\phi}^b 
                \partial_k \hat{\phi}^c \\
\pnu \tilde{F}^{o\nu}&= \frac{1}{2}\epsilon^{o\nu\sigma\rho} \pnu F^{\sigma\rho} \\ 
&=    \frac 12\epsilon^{\nu\sigma\rho} \pnu F^{\sigma\rho}\ \ 
                    \mbox{for a fix} \ \ \mu = 0 
\end{align*}
where $\nu, \ \sigma, \ \rho$ \ are    variables \cite{14}.
$$
  B_i = \frac 12 \epsilon_{ijk} F^{jk} \ \ \mbox{and} \ \
                 B^i = -\frac 12 \epsilon^{ijk} F_{jk}, 
$$
or \ $\ds B_{\nu} = \dfrac 12 \epsilon_{\nu\sigma\rho} F^{\sigma\rho}$ \ 
or \ $B^{\nu} = -\dfrac 12 \epsilon^{\nu\sigma\rho} F_{\sigma\rho}$.
\begin{align*}
\nabla \cdot \vec{B} &= \partial_{\nu^{1}}B_{\nu^{1}}  
               = \partial_i B_i = \pnu B_{\nu}\\
\nabla\cdot\vec{B} &= \frac 12 \epsilon_{\nu\sigma\rho} \pnu F^{\sigma\rho}
                        \ \   \mbox{. Clearly, we get} \ \ \nabla\cdot\vec{B}  = K^o, \\
M &= \frac{1}{4\pi} \int \nabla\cdot\vec{B}\  d^3x \ \ \mbox{so that} \ \
\nabla\cdot\vec{B} = 4\pi g \ \delta^3(x).
\end{align*}
Therefore
$M = \dfrac{1}{4\pi} \ds\int 4\pi g\ \delta^3(x)\  d^3x =
          g \underbrace{\ds\int \delta^3(x)\ d^3x}_{
       \ =  \ 1 \ \mbox{Dirac Delta}} \ \ $ 
       \newline
       so we can simply conclude that
\begin{equation}
M = g  
\end{equation}\label{eq8}
\begin{align*}
Q &= \frac{1}{4\pi} \int dS_{a} \phi^{a} = \frac{1}{8\pi} 
        \int  \epsilon^{\mu\nu} \epsilon^{abc} \phi^{a}\partial_{\mu} 
        \phi^{b}\partial_{\nu}\phi^{c} dS_{j} \\ 
Q &= \left( \int d^{3}x K_{o}\right) \times \  \mbox{const.},\ \ 
        \mbox{const} = \frac{1}{4\pi e} \ \ \mbox{and}\ \ 
        dS_{a} = \frac{1}{2} \epsilon^{\mu\nu} \epsilon^{abc} 
        \frac{\partial x_{b}}{\partial \sigma_{\mu}} 
        \frac{\partial x_{c}}{\partial \sigma_{\nu}} d^{2}\sigma \\ 
Q &= \frac{1}{8\pi} \int  \epsilon_{ijk} \epsilon^{abc} 
        \partial_{i}\hat{\phi}^{a}\partial_{j}\hat{\phi}^{b}
        \partial_{k}\hat{\phi}^{c} d^{3}x 
\end{align*}
\begin{align*}
Q &= \frac{1}{8\pi} \int  \epsilon_{ijk} \epsilon^{abc} 
        \partial_{i}(\hat{\phi }^{a}\partial _{j}\hat{\phi }^{b}
        \partial_{k}\hat{\phi }^{c}) d^{3}x \\ 
Q &= \frac{1}{8\pi} \int  \epsilon_{jk} \epsilon^{abc} \hat{\phi}^{a}
        \partial_{j}\hat{\phi }^{b}\partial_{k}\hat{\phi }^{c} d^{2}x \\ 
Q &= \frac{1}{8\pi} \int  \epsilon_{jk} \epsilon^{abc} \hat{\phi}^{a} 
        \frac{\partial \hat{\phi}^{b}}{\partial x^{j}}
        \frac{\partial \hat{\phi}^{c}}{\partial x^{k}} d^{2}x \\ 
Q &= \frac{1}{8\pi} \int  \epsilon_{jk} \hat{\phi}\cdot (\partial_{j}
        \hat{\phi}\times\partial_{k}\hat{\phi}) d^{2}x \\ 
Q &= \frac{1}{8\pi} \int  \epsilon_{\mu\nu} \hat{\phi}\cdot (\partial_{\mu}
        \hat{\phi }\times\partial_{\nu}\hat{\phi}) d^{2}x 
\end{align*}
since $\hat{\phi} = \dfrac{\phi}{|\phi |} $  therefore
\begin{align*}
F_{\mu\nu} &= \partial_{\mu}A_{\nu} - \partial_{\nu}A_{\mu} - 
        \frac{1}{e} \epsilon_{abc}\hat{\phi}^{a}(\partial_{\mu}
        \hat{\phi}^{b})(\partial _{\nu }\hat{\phi }^{c}) \\ 
B^{i} &= -B_{i} = -\frac{1}{2} \epsilon ^{ijk} F_{jk} \mbox{ and we get, } \\ 
B_{i} &= \frac{1}{2} \epsilon_{ijk} F^{jk} \\ 
\nabla \cdot \vec{B} &= \partial_{i}B_{i} = \frac 12 \epsilon_{ijk} 
        \partial_{i} F_{jk} \\ 
\nabla \cdot \vec{B} &= \partial_{i}B_{i} = \frac 12 \epsilon_{ijk} 
        \partial_{i} \Big[(\partial_{j}A_{k} - \partial_{k}A_{j}) - 
        \frac{1}{e} \epsilon_{abc}\hat{\phi}^{a}\partial_{j}\hat{\phi}^{b}
        \partial_{k}\hat{\phi }^{c}\Big]\\
F_{jk} &= \partial _{j}A_{k} - \partial _{k}A_{j} - \dfrac 1e 
\epsilon _{abc} \hat{\phi }^{a}\partial _{j}\hat{\phi }^{b}\partial 
_{k}\hat{\phi }^{c}.
\end{align*}
\begin{align*}
&\partial _{i}(\partial _{j}A_{k} - \partial _{k}A_{j}) = 0 
\mbox{. This gives,} \\
\partial _{i}F_{jk} = &-\frac{1}{2e} \epsilon _{abc} \partial 
_{i}\hat{\phi }^{a}\partial _{j}\hat{\phi }^{b}\partial _{k}\hat{\phi }^{c} 
\mbox{ to get, } \\ 
\nabla \cdot \vec{B} = &\ \partial _{i}B_{i} = -\frac{1}{2e} \epsilon 
_{abc} \epsilon _{ijk} \partial _{i}\hat{\phi }^{a}\partial 
_{j}\hat{\phi }^{b}\partial _{k}\hat{\phi }^{c} 
\end{align*}
But $\nabla \cdot \vec{B} = K^{o} $ so that 
$$
K^{o} = -\frac{1}{2e} \epsilon _{abc} \epsilon ^{ijk} \partial 
_{i}\hat{\phi }^{a}\partial _{j}\hat{\phi }^{b}\partial _{k}\hat{\phi }^{c}
$$
The above is the derivation of equation (1).  

\medskip

If we compare the expression for $Q$ and $M$ we find that

\medskip

\noindent \qquad \qquad\qquad \qquad
$\ds Q = \dfrac Me$

\medskip

\noindent \qquad \qquad\qquad \qquad
$\ds M = \dfrac{1}{4\pi } \int  K^{o} d^{3}x = \dfrac{1}{4\pi } 
\int \nabla \cdot \vec{B}d^{3}x$

\medskip

\noindent \qquad \qquad\qquad \qquad
$\nabla \cdot \vec{B} = 4\pi g \delta ^{3}(x)$ 

\medskip

\noindent therefore 
$M = \dfrac{1}{4\pi} \ds\int 4\pi g \delta^{3}(x) d^{3}x = 
        g\int \delta^{3}(x) d^{3}x$.

\medskip
\noindent Thus $M = g$ since $\ds\int \delta^{3}(x) d^{3}x = 1$ (Dirac Delta), 
thus,
\begin{equation}\label{eq9}
    Q = \frac ge 
\end{equation}
\noindent where $g$ is the magnetic strength, 
\noindent and, according to (Dirac Quantization Condition)  
$g = \dfrac{\hbar c}{2e} n$, $n = 0$, $\pm 1,\ldots$

\medskip

\noindent therefore
$Q=\dfrac{\hbar c}{2e^{2}}n$, $n = 0$, $\pm 1,\ldots$
\noindent Thus, the topological charge $Q$ is quanitized.

\bigskip
\bigskip
Let us next prove the Dirac Quantization Condition is derived from the 
Klein-Gordon equation.\\
\bigskip
\noindent {\bf Proof:}  The solution for Klein-Gordon (K-G) equation is
$$
\Psi   = |\Psi | \exp \frac{i}{\hbar} (\vec{P}\cdot\vec{r} - Et),
$$
but for the time independent equation, 
$\Psi   = |\Psi | \exp \dfrac{i}{\hbar} (\vec{P}\cdot\vec{r})$.

In presence of an electromagnetic field $\vec{P} \rightarrow \vec{P} - 
\frac{e}{c} \vec{A}$, and, due to the influence
of magnetic monopole we will get \cite{12}\cite{15}
\begin{align*}
\Psi &\rightarrow |\Psi|\exp\frac{i}{\hbar}
    \vec{P}\cdot\vec{r} 
    \exp\frac{-ie}{\hbar c} \vec{A}\cdot\vec{r} \\
\Psi &\rightarrow |\Psi|\exp\frac{i}{\hbar}
    \vec{P}\cdot\vec{r} 
    \exp\frac{-ie}{\hbar c} \oint \vec{A}\cdot d\vec{\ell}
\end{align*}
where  $\vec{A}\cdot\vec{r} \rightarrow \oint 
\vec{A}\cdot d\vec{\ell}$
close cycle for a periodic function.  Therefore
$$
\Psi  \rightarrow  \Psi  \exp \frac{-ie}{c\hbar} 
           \oint \vec{A}\cdot d\vec{\ell} \quad 
\mbox{so that} \quad \exp  \frac{-ie}{c\hbar}
\oint \vec{A}\cdot d\vec{\ell} = 1,
$$
$$
  e^{\frac{-ie}{c\hbar} \oint \vec{A}\cdot d\vec{\ell}} = 1 \quad 
\mbox{then} \quad \frac{-ie}{c\hbar}\oint \vec{A}\cdot d\vec{\ell}
= -2\pi ni
$$
changing from line integral to surface integral we have
$$
\oint \vec{A}\cdot d\vec{\ell} = \int\nabla \times\vec{A}\cdot d\vec{S}, 
 \quad \vec{B} = \nabla \times\vec{A} \mbox{ gives}
$$
$$
    \oint \vec{A}\cdot d\vec{\ell} = \int \vec{B}\cdot d\vec{S}
$$
changing from surface integral to volume integral
$$
\int \vec{B}\cdot d\vec{S} = \int \nabla \cdot \vec{B} \ d^{3}x
$$

\medskip

\noindent to get \qquad 
$\ds\oint \vec{A}\cdot d\vec{\ell} = 
\ds\int \nabla \cdot \vec{B} \ d^{3}x$

\medskip

\noindent or  
$
    \dfrac{e}{c\hbar} \ds\oint \vec{A}\cdot d\vec{\ell} 
  = \ds\dfrac{e}{c\hbar} \ds\int \nabla \cdot \vec{B} \ d^{3}x,
$
\medskip

\noindent therefore
$    
      2\pi n = \dfrac{e}{c\hbar} \ds\int \nabla\cdot \vec{B} d^3x
$.
\medskip

\noindent But  
$\nabla \cdot \vec{B} = 4\pi g \delta^3(x) \ $gives$ \ 
    2\pi n = \dfrac{e}{c\hbar} \ds\int 4\pi g \delta^3(x) \ d^3x$.

\medskip

\noindent thus \qquad 
$\ds n = \frac{2e}{c\hbar}\ g \int \delta^3(x) \ d^3x$.

\medskip

\noindent But $\ds\int \delta^3(x) d^3x = 1, \   \mbox{Dirac Delta, whence}$
$$
    n = \frac{2eg}{c\hbar} \quad \mbox{therefore} \quad g = 
\frac{\hbar cn}{2e}, \quad \mbox{Dirac Quantization Condition}
$$
\\

The result thus far applies to bosons, because the Dirac quanitization condition is derived from the Klein-Gordon equation, which describes only spin-zero particles (bosons).  We will now proceed to address the fermion case.

\newpage


\noindent
{\bf Discussion on the Quantization of \\ 
Topological Charge for Fermions}\\
\\

To prove that each of the four components  $\Psi_i$ of Dirac's equation 
satisfies the Klein-Gordon equation.

\bigskip

\noindent {\bf Proof:}  
The covariant Dirac equation is given by \cite{15}
$$
  (i \gamma^{\mu}\partial_{\mu} - m) \Psi = 0 \quad 
\mbox{or} \quad i \gamma^{\mu}\partial_{\mu} \Psi = m\Psi.
$$

Operate on the covariant Dirac equation by 
$\gamma^{\nu}\partial_{\nu}, $ to get:
$$
\gamma^{\nu}\partial_{\nu} (i \gamma^{\mu}
\partial_{\mu} - m) \Psi = 0, or
$$
$$(i \gamma^{\nu}\gamma^{\mu}
\partial_{\nu}\partial_{\mu} - m \gamma^{\nu}\partial_{\nu}) \Psi = 0
$$
but
$$
\gamma^{\nu}\gamma^{\mu}\partial_{\nu}\partial_{\mu} = \sum_{\mu\nu}
\gamma^{\nu}\gamma^{\mu} \pnu\pmu= \sum_{\nu,\mu} (\frac 12
\gamma^{\nu}\gamma^{\mu} + \gamma^{\mu}\gamma^{\nu}) \partial_{\nu}\partial_{\mu}
$$
so that
$$
\qquad  i \frac 12 
(\gamma^{\nu}\gamma^{\mu} + \gamma^{\mu}\gamma^{\nu})
\partial_{\nu}\partial_{\mu} \Psi - m \gamma^{\nu} \partial_{\nu}\Psi = 0
$$
By the anticommutation relation: 
$\gamma^{\mu}\gamma^{\nu} + \gamma^{\nu}\gamma^{\mu} =  2g^{\mu\nu}$ \\ 
the preceding equation becomes

$$
ig^{\mu\nu} \partial_{\nu}\partial_{\mu} \Psi - m 
\underbrace{\gamma^{\nu}\partial_{\nu} \Psi}_{\frac{m\Psi}{i}} = 0
$$

Note that the covariant Dirac equation will result in $i\gamma^{\nu}\partial_{\nu} 
\Psi = m\Psi$ or $\gamma^{\nu}\partial_{\nu}\Psi = \dfrac{m\Psi}{i}$.

\medskip

\noindent  so that, \quad $ig^{\mu\nu} 
\partial_{\nu}\partial_{\mu} \Psi - m 
\ds\Big(\dfrac{m\Psi}{i}\Big) = 0$

\medskip

\noindent $g^{\mu\nu} \partial_{\nu}\partial_{\mu} = \partial^{\mu}
\partial_{\mu}$ \  
where \ $g^{\mu\nu}$ is the metric tensor, therefore $i 
\partial^{\mu}\partial_{\mu} \Psi - m (\dfrac{m\Psi}{i}) = 0,    \quad i = 
-\dfrac{1}{i}$
$$
i \partial^{\mu}\partial_{\mu} \Psi + m^2 \Psi = 0 \quad or 
\quad 
i(\partial^{\mu}\partial_{\mu} + m^2) \Psi = 0.
$$
But since $\partial^{\mu}\partial_{\mu} = \Box^2 \quad $then \quad 
$(\Box^2 + m^2) \Psi = 0$, or 
$(\Box^2 + m^2) \Psi_i = 0$, therefore each of the four components 
$\Psi_i$ satisfies the Klein-Gordon equation. However in the presence of 
electromagentic field that may not be the case.

\def\theequation{\arabic{equation}}

\qquad In the presence of electromagnetic field, Dirac's equation in flat spacetime is \cite{17}
$$
   \left[ - (-\frac 1i \frac{\partial}{\partial t} + e\phi)^2
   + (\frac 1i \nabla + e\vec{A})^2 + m^2 + \vec{\sigma} \cdot \vec{B}
   - ie \rho_1(\vec{\sigma}\cdot \vec{E})\right] \Psi (\vec{x},t) = 0
$$
$\vec{\sigma}\cdot \vec{B} - ie \rho_1(\vec{\sigma}\cdot \vec{E}) =$ function
of the electromagnetic field $(\vec{A}({x}),i\phi(t))$ \ so that
$$
    \vec{\sigma}\cdot \vec{B} - ie\rho_1(\vec{\sigma}\cdot \vec{E}) 
   = f(\vec{x},t) = f(x,t)
$$
Therefore
\begin{equation}\label{eq16}
   \left[ - (-\frac 1i \frac{\partial}{\partial t} + e\phi)^2
   + (\frac 1i \nabla + e\vec{A})^2 + m^2 +  f({x},t)\right]
     \Psi (\vec{x},t) = 0.
\end{equation}

Klein-Gordon's equation in flat spacetime in the presence of electromagnetic
field is \cite{17}:
$$
   \left[(-\frac 1i \frac{\partial}{\partial t} + e\phi)^2
   - (\frac 1i \nabla + e\vec{A})^2\right] \Psi (\vec{x},t) = m^2
    \Psi (\vec{x},t)
$$
$$
    (E + e\phi)^2 - (\vec{P} + e\vec{A})^2 = m^2
$$
\noindent The above two equations lead us to
$$
 \left[(-\frac 1i \frac{\partial}{\partial t} + e\phi)^2
   - (\frac 1i \nabla + e\vec{A})^2\right] \Psi (\vec{x},t) 
   = \left[(E + e\phi)^2 - (\vec{P} + e\vec{A})^2\right]\Psi (\vec{x},t)
$$
Thus the time independent equation is:
$$
   (\frac 1i \nabla + e\vec{A})^2
   \Psi (\vec{x}) = (\vec{P} + e\vec{A})^2 \Psi (\vec{x})
$$
and the time dependent equation is:
$$
   (-\frac 1i \frac{\partial}{\partial t} + e\phi)^2
   \Psi (t) = (E + e\phi)^2 \Psi (t)
$$
Equation (5) can be written as
$$
   \left[(\frac 1i \nabla + e\vec{A})^2
   - (-\frac 1i \frac{\partial}{\partial t} + e\phi)^2 + m^2\right]
     \Psi (\vec{x},t) =  -f(x,t) \Psi (\vec{x},t).
$$
This is regarded as nonhomogeneous equation whose solution is
$$
  \Psi (\vec{x},t) =  \phi(\vec{x},t) + \int G_p(\vec{x},\vec{x}^{\prime})
   [-f(x',t)] \Psi (\vec{x}^{\prime},t) d^3x'
$$
where $ \phi(\vec{x},t)$ is the solution of the homogeneous equation.
$$
  \left[(-\frac 1i \nabla + e\vec{A})^2
   - (-\frac 1i \frac{\partial}{\partial t} + e\phi)^2\right]
     \phi(\vec{x},t) =  0
$$
is the homogeneous equation which is Klein-Gordon's equation representing
a particle in an electromagnetic field.  The time independent equation
corresponding to the above equation is
$$
   (\frac 1i \nabla + e\vec{A})^2 \Psi (x) = (\vec{P} + e\vec{A})^2 \Psi (x)
$$
there will be a restriction on the above equation due to the 
presence of magnetic monopole.

One restriction on $\vec{A}$ is that it must be singular along the
negative $z$-axis, there must be an unphysical singularity (Dirac
String).  Such a descriptive corresponds only to a wave function in
a presence of magnetic monopole which can be obtained by making the
standard substitution $\vec{P} \ra \vec{P} + e\vec{A}$, and also
demanding the wave function to be single valued when going around 
the loop, i.e.,
$$
   \phi(\vec{P} \cdot \vec{x} + 2\pi n) = \phi(\vec{P} \cdot \vec{x}),
$$
although $\vec{P} \cdot \vec{x} \neq \vec{P} \cdot \vec{x} + 2\pi n$
so that, the line integral around the Dirac String must be $2\pi n$,
$n =$ integer.  This condition will give rise to $\exp + ie \vec{A}
\cdot \vec{x} = 1$ so that, $2\pi n = e\ds\oint \vec{A} \cdot d\vec{\ell}$.
Under these conditions we examine the equation
$$
   (\frac 1i \nabla + e\vec{A})^2 \phi(\vec{x}) = (\vec{P} + e\vec{A})^2 
    \phi(\vec{x})
$$
to find that the only way these conditions can be satisfied if
$$   (\frac 1i \nabla + e\vec{A})^2\rightarrow (\frac 1i \nabla')^2
   \quad \mbox{and} \quad \phi(\vec{x}) \rightarrow \phi'(\vec{x}^{\prime})
$$
let $e$ in the above equation be replaced by $-e$ to get
$$
   (\frac 1i \nabla - e\vec{A})^2\rightarrow (\frac 1i \nabla')^2
   \quad \mbox{and} \quad \phi(\vec{x}) \rightarrow \phi'(\vec{x}^{\prime})
$$
so that if $\phi'(\vec{x}^{\prime}) \equiv \phi(\vec{P} \cdot \vec{x} + 2\pi n)$
then $\phi'(\vec{x}^{\prime}) = \phi(\vec{P} \cdot \vec{x})$ or $\phi'(\vec{x}^
{\prime})
= \phi(\vec{x}) \equiv \phi(\vec{P} \cdot \vec{x})$, where $\phi(\vec{P} \cdot
\vec{x}) =  \phi(\vec{P} \cdot\vec{x} + 2\pi n)$.

If $\phi(\vec{P} \cdot\vec{x}) = \phi\big[(\vec{P} - e\vec{A})\cdot \vec{x}\big]
$ which implies that $ e\vec{A} \cdot \vec{x} = 2\pi n$, we can clearly see that the only way these
conditions are met if $(\dfrac 1i \nabla - e\vec{A})^2\rightarrow 
(\dfrac 1i \nabla')^2$ and $\phi(\vec{x}) \rightarrow \phi'(\vec{x}^{\prime})$
so that $(\dfrac 1i \nabla')^2\phi'(\vec{x}^{\prime}) = (\vec{P} - e\vec{A})
\phi'(\vec{x}^{\prime})$ which gives $\phi'(\vec{x}^{\prime}) = A_1 e^{i(\vec{P} 
- e\vec{A}) \cdot \vec{x}^{'}}$ that satisfies the above equations and conditions,
therefore $\phi (\vec{x},t) = \phi(t) e^{i(\vec{P} - e\vec{A}) \cdot \vec{x}}$.

\medskip
\noindent This leads us to \quad $\Psi (\vec{x},t) = \phi(t) 
e^{i(\vec{P} - e\vec{A}) \cdot \vec{x}} 
+ \ds\int G_p(\vec{x},\vec{x}^{\prime},t)[-f(x',t)]\Psi (\vec{x}^{\prime},t)
d^3x'$

\noindent or \quad  $\Psi (\vec{x},t) = \phi(t)e^{i(\vec{P} - e\vec{A}) \cdot \vec{x}}
-  \ds\int G_p(\vec{x},\vec{x}^{\prime},t)f(x',t)\Psi (\vec{x}^{\prime},t)
d^3x'$.

In the absence of electromagnetic field and of magnetic monopole we have seen above
 that each of the four components of Dirac's Matrix satisfies
the Klein-Gordon equation, therefore
$$
    \Psi (\vec{x},t) = A_1e^{i(\vec{P} \cdot \vec{x} - Et)} = A_1e^{i\vec{P}
 \cdot \vec{x}} e^{-Et}.
$$
For $\Psi (\vec{x},t)$ to be single valued when $\vec{A}$ is singular 
we should have
\begin{align*}
  A_1e^{i\vec{P} \cdot \vec{x}} e^{-iEt} &= \phi(t)e^{i\vec{P}
 \cdot \vec{x}} e^{-ie\vec{A}\cdot \vec{x}} - 
\int G_p(\vec{x},\vec{x}^{\prime},t)f(x',t)\Psi (\vec{x}^{\prime},t)d^3x'\\
e^{-ie\vec{A}\cdot \vec{x}} & = A_1\phi^{-1}(t)e^{-iEt} + \phi^{-1}(t)
e^{-i\vec{P} \cdot \vec{x}}
\int G_p(\vec{x},\vec{x}^{\prime},t)f(x',t)\Psi (\vec{x}^{\prime},t)d^3x'
\end{align*}
in order for the quantization to take place.

We have seen already that $e^{-ie\vec{A}\cdot\vec{x}}$  (where $x$ here is equivalent to $r$ as used earlier) should
have been equal to a constant, when the wave function is single valued.
However, we can see from the above equation that $e^{-ie\vec{A}\cdot\vec{x}}
=$ function of $x \neq$ constant.  Thus, we conclude that, in a case of
fermions the topological charge is not quantized.

ACKNOWLEDGEMENT:

The author would like to thank Alan Chodos, Kevin Sharpe, and Charles Hoang for fruitful discussion.

\bibliographystyle{amsplain}

\end{document}